\documentclass[12pt,twoside,a4paper]{article}
\usepackage{graphicx}
\usepackage[dvips]{epsfig}
\usepackage{amssymb}
\usepackage{rotating}
\usepackage{a4wide}

\def\gsim{\mathrel{\rlap{\lower4pt\hbox{\hskip1pt$\sim$}}\raise1pt\hbox{$>$}}}

\def\Journal#1#2#3#4{{#1} {\bf #2}, #3 (#4)}

\def\PRL{\em Phys. Rev. Lett.}

\newcommand{\be}{\begin{equation}}
\newcommand{\ee}{\end{equation}}
\newcommand{\ba}{\begin{eqnarray}}
\newcommand{\ea}{\end{eqnarray}}

\def\gsim{\mathrel{\rlap{\lower4pt\hbox{\hskip1pt$\sim$}}\raise1pt\hbox{$>$}}}
\def\lsim{\mathrel{\rlap{\lower4pt\hbox{\hskip1pt$\sim$}}\raise1pt\hbox{$<$}}}

\def\gsim{\mathrel{\rlap{\lower4pt\hbox{\hskip1pt$\sim$}}\raise1pt\hbox{$>$}}}

\begin{document}
\setcounter{secnumdepth}{4}
\setcounter{tocdepth}{4}

%%%%%%%%%%%%%%%%%%%%%%%%%%%%%%%%%%%%%%%%%%%%%%%%%%%%%%%%%%%%%%%%%%%%%%%%%%%%%
%%%%%%%%%%%%%%%%%%%%%%%%%%%%%%%%%%%%%%%%%%%%%%%%%%%%%%%%%%%%%%%%%%%%%%%%%%%%%
\vfill
\begin{center} \LARGE Letter--of--Intent \\ for \end{center}
\thispagestyle{empty}
\vfill
\begin{center}
  {\noindent\Huge Measurement of the Spin--Dependence of the
  $\bar{p}p$ Interaction at the AD--Ring}
\end{center}
\begin{center} \LARGE (${\cal PAX}$ Collaboration) \end{center}
\vfill
\begin{center} {\noindent\Large J\"ulich, November 2005\\} \end{center}
\vfill
\cleardoublepage

%\vfill
\begin{center} \LARGE   Letter--of--Intent\\ for \end{center}
%\vfill
\begin{center}
  {\noindent\Huge Measurement of the Spin--Dependence of the
  $\bar{p}p$ Interaction at the AD--Ring}
\end{center}
\begin{center} \LARGE (${\cal PAX}$ Collaboration) \end{center}
%\vfill 

%%%%%%%%%%%%%%%%%%%%%%%%%%%%%%%%%%%%%%%%%%%%%%%%%%%%%%%%%%%%%%%%%%%%%%%%%%%%%
\begin{abstract}
An internal polarized hydrogen storage cell gas target is proposed for
the AD--ring to determine for the first time the two total
spin--dependent cross sections $\sigma_1$ and $\sigma_2$ at antiproton
beam energies in the range from 50 to 200 MeV. The data will allow the
definition of the optimum working parameters of a dedicated Antiproton
Polarizer Ring (APR), which has recently been proposed by the PAX
collaboration for the new Facility for Antiproton and Ion Research
(FAIR) at GSI in Darmstadt, Germany. The availability of an intense
beam of polarized antiprotons will provide access to a wealth of
single-- and double--spin observables, thereby opening a new window to
QCD transverse spin physics.  The physics program proposed by the PAX
collaboration includes a first measurement of the transversity
distribution of the valence quarks in the proton, a test of the
predicted opposite sign of the Sivers--function, related to the quark
distribution inside a transversely polarized nucleon, in Drell--Yan
(DY) as compared to semi--inclusive Deep Inelastic Scattering, and a
first measurement of the moduli and the relative phase of the
time--like electric and magnetic form factors $G_{E,M}$ of the proton.
\end{abstract}
%%%%%%%%%%%%%%%%%%%%%%%%%%%%%%%%%%%%%%%%%%%%%%%%%%%%%%%%%%%%%%%%%%%%%%%%%%%%%
\begin{center}
\section*{Spokespersons:} 
\end{center}
\begin{center}
Paolo Lenisa\\
Istituto Nazionale di Fisica Nucleare, Ferrara, Italy \\
E--Mail: lenisa@fe.infn.it\\ 
\end{center}

\begin{center}
Frank Rathmann\\ 
Institut f\"ur Kernphysik, Forschungszentrum J\"ulich, Germany \\
E--Mail: f.rathmann@fz--juelich.de
\end{center}

\clearpage
\begin{center}
\section*{Institutions\footnote{A complete list of PAX
collaborators is given in Appendix A.}} 
\small
Yerevan Physics Institute, Yerevan, Armenia\\
Department of Subatomic and Radiation Physics, University of Gent, Gent, Belgium\\
Department of Modern Physics, University of Science and Technology of China, Hefei, China\\
School of Physics, Peking University, Beijing, China\\
Centre de Physique Theorique, Ecole Polytechnique, Palaiseau, France\\
Institute of High Energy Physics and Informatization, Tbilisi State University, Tbilisi, Georgia\\
Nuclear Physics Department, Tbilisi State University, Tbilisi, Georgia\\
Forschungszentrum J\"ulich, Institut f\"ur Kernphysik, J\"ulich, Germany\\
Helmholtz--Institut f\"ur Strahlen-- und Kernphysik, Universit\"at Bonn, Bonn, Germany\\
Institut f\"ur Theoretische Physik II, Ruhr Universit\"at Bochum, Bochum, Germany\\
Physikalisches Institut, Universit\"at Er\-langen--N\"urn\-berg, Erlangen, Germany\\
Unternehmensberatung und Service--B\"uro (USB), Gerlinde Schulteis \& Partner GbR, Langenbernsdorf, Germany\\
School  of Mathematics, Trinity College, University of Dublin, Dublin, Ireland\\
Dipartimento di Fisica Teorica, Universita di Torino and INFN, Torino, Italy\\
Dipartimento di Fisica, Universita$'$ di Cagliari and INFN, Cagliari, Italy\\
Dipartimento di Fisica, Universita$'$ di Lecce and INFN, Lecce, Italy\\
Istituto Nazionale di Fisica Nucleare, Ferrara, Italy\\
Istituto Nazionale di Fisica Nucleare, Frascati, Italy\\
Universita$'$ del Piemonte Orientale $''$A. Avogadro$''$ and INFN, Alessandria, Italy\\
Universita' dell'Insubria, Como and INFN sez., Milano, Italy\\
Soltan Institute for Nuclear Studies, Warsaw, Poland\\
Budker Institute for Nuclear Physics, Novosibirsk, Russia\\
Joint Institute for Nuclear Research, Dubna, Russia\\
Petersburg Nuclear Physics Institute, Gatchina, Russia\\
Institute of High Energy Physics, Protvino, Russia\\
Institute for Theoretical and Experimental Physics, Moscow, Russia\\
Lebedev Physical Institute, Moscow, Russia\\
Physics Department, Moscow Engineering Physics Institute, Moscow, Russia\\
Institute of Experimental Physics, Slovak Academy of Sciences and P.J. Safarik University, Faculty of Science, Kosice, Slovakia\\
Department of Radiation Sciences, Nuclear Physics Division, Uppsala, Sweden\\
Collider--Accelerator Department, Brookhaven National Laboratory, Brookhaven, USA\\
Department of Physics, University of Virginia, Virginia, USA\\
RIKEN BNL Research Center, Brookhaven National Laboratory, Brookhaven, USA\\
University of Wisconsin, Madison, USA\\
\end{center}
%%%%%%%%%%%%%%%%%%%%%%%%%%%%%%%%%%%%%%%%%%%%%%%%%%%%%%%%%%%%%%%%%%%%%%%%%%%%%
\cleardoublepage

\tableofcontents

%%%%%%%%%%%%%%%%%%%%%%%%%%%%%%%%%%%%%%%%%%%%%%%%%%%%%%%%%%%%%%%%%%%%%%%%%%%%%
\cleardoublepage
\section{Introduction\label{introduction}}
\pagestyle{myheadings} \markboth{Letter--of--Intent: Spin--Dependence
of $\bar{p}p$ Interaction at the AD}{Introduction}
%%%%%%%%%%%%%%%%%%%%%%%%%%%%%%%%%%%%%%%%%%%%%%%%%%%%%%%%%%%%%%%%%%%%%%%%%%%%%
In this Letter--of--Intent, the PAX collaboration suggests to study
the polarization buildup in an antiproton beam at the AD--ring at CERN
at energies in the range from 50--200 MeV. The scientific objectives
of this experiment are twofold. The polarization buildup by spin
filtering of stored antiprotons by multiple passage through a
polarized internal hydrogen gas target gives a direct access to the
spin dependence of the antiproton--proton total cross section. Apart
from the obvious interest for the general theory of $p\bar{p}$
interactions, the knowledge of these cross sections is necessary for
the interpretation of unexpected features of the $p\bar{p}$, and other
antibaryon--baryon pairs, contained in final states in $J/\Psi$ and
$B$--decays. Simultaneously, the confirmation of the polarization
buildup of antiprotons would pave the way to high--luminosity
double--polarized antiproton--proton colliders, which would provide a
unique access to transverse spin physics in the hard QCD regime. Such
a collider has been proposed recently by the PAX Collaboration
\cite{PAX-TP} for the new Facility for Antiproton and Ion Research
(FAIR) at GSI in Darmstadt, Germany, aiming at luminosities of
10$^{31}$~cm$^{-2}$s$^{-1}$. An integral part of such a machine is a
dedicated large--acceptance Antiproton Polarizer Ring (APR).

Here we recall, that for more than two decades, physicists have tried
to produce beams of polarized antiprotons \cite{krisch}, generally
without success.  Conventional methods like atomic beam sources (ABS),
appropriate for the production of polarized protons and heavy ions
cannot be applied, since antiprotons annihilate with matter.
Polarized antiprotons have been produced from the decay in flight of
$\bar{\Lambda}$ hyperons at Fermilab. The intensities achieved with
antiproton polarizations $P>0.35$ never exceeded $1.5 \cdot
10^5$~s$^{-1}$ \cite{grosnick}.  Scattering of antiprotons off a
liquid hydrogen target could yield polarizations of $P\approx 0.2$,
with beam intensities of up to $2 \cdot 10^3$~s$^{-1}$ \cite{spinka}.
Unfortunately, both approaches do not allow efficient accumulation in
a storage ring, which would greatly enhance the luminosity.  Spin
splitting using the Stern--Gerlach separation of the given magnetic
substates in a stored antiproton beam was proposed in 1985
\cite{niinikoski}.  Although the theoretical understanding has much
improved since then \cite{cameron}, spin splitting using a stored beam
has yet to be observed experimentally. In contrast to that, a
convincing proof of the spin--filtering principle has been produced by
the FILTEX experiment at the TSR--ring in Heidelberg \cite{TSR}.

The experimental basis for predicting the polarization buildup in a
stored antiproton beam is practically non--existent. The AD--ring at
CERN is a unique facility at which stored antiprotons in the
appropriate energy range are available and whose characteristics meet
the requirements for the first ever antiproton polarization buildup
studies.  Therefore, it is of highest priority for the PAX
collaboration to perform spin filtering experiments using stored
antiprotons at the AD--ring of CERN. Once the experimental data base
will be made available by the AD experiments, the final design of a
dedicated APR can be targeted.  In addition, a few dedicated spin
filtering experiments carried out with protons at the Cooler
Synchroton COSY at J\"ulich, Germany, will enhance our general
understanding of these processes and allow us to commission the
additional equipment needed for the spin filtering experiments at the
AD.

%%%%%%%%%%%%%%%%%%%%%%%%%%%%%%%%%%%%%%%%%%%%%%%%%%%%%%%%%%%%%%%%%%%%%%%%%%%%%
\section{Physics Case \label{physicscase}}
\pagestyle{myheadings} \markboth{Letter--of--Intent: Spin--Dependence
of $\bar{p}p$ Interaction at the AD}{Physics Case}
%%%%%%%%%%%%%%%%%%%%%%%%%%%%%%%%%%%%%%%%%%%%%%%%%%%%%%%%%%%%%%%%%%%%%%%%%%%%%
\subsection{$N\bar{N}$ Double-Spin Observables from Spin Filtering}
%%%%%%%%%%%%%%%%%%%%%%%%%%%%%%%%%%%%%%%%%%%%%%%%%%%%%%%%%%%%%%%%%%%%%%%%%%%%%
The two double--spin observables, which can be measured by the
spin--filtering technique, are the spin--dependent cross sections
$\sigma_1$ and $\sigma_2$ in the parameterization of the total
hadronic cross section $\sigma_{\mathrm{tot}}$ \cite{bilenky}, written
as
\begin{eqnarray}
\sigma_{\mathrm{tot}}=\sigma_0 + \sigma_1 (\vec{P} \cdot \vec{Q}) +
\sigma_2 (\vec{P} \cdot \hat{k})(\vec{Q}\cdot \hat{k})\;,
\end{eqnarray}
where $\sigma_0$ denotes the total spin--independent hadronic cross
section, $\sigma_1$ the total spin--dependent cross section for
transverse orientation of beam polarization $P$ and target
polarization $Q$, $\sigma_2$ denotes the total spin--dependent cross
section for longitudinal orientation of beam and target
polarizations. (Here we use the nomenclature introduced by Bystricky,
Lehar, and Winternitz \cite{bystricky}, where $\hat{k}=\vec{k}/
|\vec{k}|$ is the unit vector along the collision axis.) Such
observables would improve substantially the modern phenomenology of
proton--antiproton interactions based on the experimental data
gathered at LEAR (for a review and references, see \cite{klempt}).

\begin{figure}[hbt]
 \begin{center}
\includegraphics[width=0.5\linewidth]{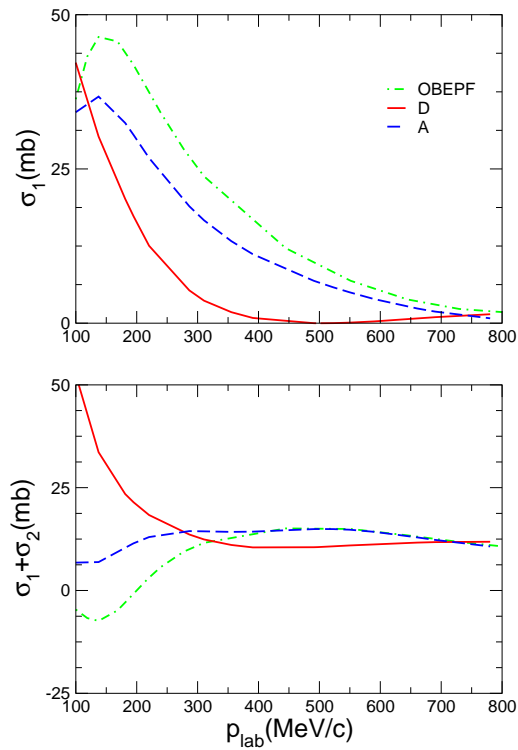}
  \parbox{14cm}{\caption{\label{pbarpmodels}\small The spin--dependent
cross sections $\sigma_1$ and $\sigma_1+\sigma_2$, cf. Eq. (1), as
predicted by the $N\bar N$ models D \cite{Mull}, A \cite{Hippchen} and
OBEPF \cite{Haidenbauer} of the J\"ulich group.
}}
\end{center}
\end{figure}

The suggested spin--filtering experiment at the AD of CERN constitutes
a unique opportunity to measure for the first time these observables
in the 50--200 MeV energy range. The measurements of $\sigma_1$ and
$\sigma_2$ will be carried out in the transmission mode. The
separation of the elastic scattering and annihilation contributions to
$\sigma_1$ and $\sigma_2$ requires the integration of the
double--polarized elastic cross section over the full angular
range. Although such measurements seem not feasible with the
anticipated luminosity using the HERMES internal polarized target
installed at the AD, the obtained results on $\sigma_1$ and $\sigma_2$
for the total cross section would serve as an important constraint for
a new generation of baryon--antibaryon interaction models, which will
find broad application to the interpretation of the experimental data
in heavy quark physics. Regarding the main goal of the proposed
experiment -- the antiproton polarization buildup -- the expectations
from the first generation models for double--spin dependence of
$p\bar{p}$ interaction are encouraging, see Fig.~\ref{pbarpmodels}.
%---------------------------------------------------------------------------------------------------
%---------------------------------------------------------------------------------------------------
With filtering for two lifetimes of the beam, they suggest that in a
dedicated large--acceptance storage ring, antiproton beam
polarizations in the range of 15--25 \% seem achievable
\cite{heimbach}.

%%%%%%%%%%%%%%%%%%%%%%%%%%%%%%%%%%%%%%%%%%%%%%%%%%%%%%%%%%%%%%%%%%%%%%%%%%%%%
\subsection{$N\bar{N}$ Interaction from LEAR to $J/\Psi$ and $B$-decays}
%%%%%%%%%%%%%%%%%%%%%%%%%%%%%%%%%%%%%%%%%%%%%%%%%%%%%%%%%%%%%%%%%%%%%%%%%%%%%
The evidence for threshold enhancements in $B-$ and $J/\Psi$--decays
containing the baryon--antibaryon pairs -- $p\bar{p}$,
$p\bar{\Lambda}, \Lambda\bar{p}$, etc. -- was found recently at the
modern generation electron--positron colliders BES
\cite{BES1,BES2004,BES2005,BES2005-2}, BELLE
\cite{BELLE1,BELLE2,BELLE3,BELLE4} and BaBar
\cite{BaBarDecay,BaBarFF}.  These findings added to the urgency of
understanding low and intermediate energy $p\bar{p}$ interactions,
which appear to be more complex than suggested by the previous
analyses \cite{Hippchen,Mull1,Mull,Nijmegen,Paris} of the experimental
data from LEAR. The direct measurements of $\sigma_1$ and $\sigma_2$
would facilitate the understanding of the role of antibaryon--baryon
final state interactions, which are crucial for the re--interpretation
of the B--decay dynamics in terms of the Standard Model mechanisms
(see \cite{Rosner,BaBarDecay,Rosner2005} and references therein).
Especially strong theoretical activity
(\cite{Rosner,He,Bugg,Kerbikov,Sibirtsev,Ding,Datta,Yan,Chang,Chang,Loiseau,Loiseau2,Ding2}
and references therein) has been triggered by the BES finding
\cite{BES1} of the pronounced threshold enhancement in the reaction
$J/\Psi \to p\bar{p}\gamma$, including the revival of the baryonium
states \cite{Bogdanova,Shapiro} in the $p\bar{p}$ system
\cite{Datta,Yan,Chang,Chang,Loiseau,Loiseau2}. Equally important is
the recent confirmation by the BaBar collaboration \cite{BaBarFF} of
the near--threshold structure in the timelike form factor of the
proton, observed earlier at LEAR \cite{LEARFF}.  In conjunction with
the BES enhancement, the LEAR--BaBar data suggest a non--trivial
energy dependence in both the spin--singlet and spin--triplet
$p\bar{p}$ interactions, hence our special interest in $\sigma_1$ and
$\sigma_2$.

%%%%%%%%%%%%%%%%%%%%%%%%%%%%%%%%%%%%%%%%%%%%%%%%%%%%%%%%%%%%%%%%%%%%%%%%%%%%%
\subsection{Applications of Polarized Antiprotons to QCD Spin
Studies}
%%%%%%%%%%%%%%%%%%%%%%%%%%%%%%%%%%%%%%%%%%%%%%%%%%%%%%%%%%%%%%%%%%%%%%%%%%%%%
The QCD physics potential of experiments with high energy polarized
antiprotons is enormous, yet hitherto high luminosity experiments with
polarized antiprotons have been impossible. The situation could change
dramatically with the realization of spin filtering and storing of
polarized antiprotons, and the realization of a double--polarized
high--luminosity antiproton--proton collider.  The list of fundamental
physics issues for such collider includes the determination of
transversity, the quark transverse polarization inside a transversely
polarized proton, the last leading twist missing piece of the QCD
description of the partonic structure of the nucleon, which can be
directly measured only via double polarized antiproton--proton
Drell--Yan production.  Without measurements of the transversity, the
spin tomography of the proton would be ever incomplete. Other items of
great importance for the perturbative QCD description of the proton
include the phase of the timelike form factors of the proton and hard
antiproton--proton scattering.  Such an ambitious physics program has
been formulated by the PAX collaboration (Polarized Antiproton
eXperiment) and a Technical Proposal \cite{PAX-TP} has recently been
submitted to the FAIR project.  The uniqueness and the strong
scientific merits of the PAX proposal have been well received
\cite{PAX-web}, and there is an urgency to convincingly demonstrate
experimentally that a high degree of antiproton polarization could be
reached with a dedicated APR.

%%%%%%%%%%%%%%%%%%%%%%%%%%%%%%%%%%%%%%%%%%%%%%%%%%%%%%%%%%%%%%%%%%%%%%%%%%%%
\section{Measurement Technique \label{technique}}
\pagestyle{myheadings} \markboth{Letter--of--Intent: Spin--Dependence
of $\bar{p}p$ Interaction at the AD}{Measurement Technique}
%%%%%%%%%%%%%%%%%%%%%%%%%%%%%%%%%%%%%%%%%%%%%%%%%%%%%%%%%%%%%%%%%%%%%%%%%%%%%
At the core of the PAX proposal is spin filtering of stored
antiprotons by multiple passage through an internal polarized gas
target.  The feasibility of the spin filtering technique has
convincingly been demonstrated in the FILTEX experiment at TSR
\cite{TSR}: for 23 MeV stored protons, the transverse polarization
rate of $dP/dt = 0.0124 \pm 0.0006$ per hour has been reached with an
internal polarized atomic hydrogen target of areal density $6\times
10^{13}$~atoms/cm$^2$. For a proton impinging on a polarized hydrogen
gas target, the spin--dependent interaction leading to the buildup of
polarization in the beam is known; recent investigations
\cite{milstein,heimbach} have shown that an understanding and
interpretation of the FILTEX result in terms of the proton--proton
interaction is available.

The polarization buildup of the beam as a function of filter time $t$
can be expressed in the absence of depolarization as \cite{TSR}
\begin{eqnarray}
P(t)=\tanh(t/\tau_1)
\end{eqnarray}
The time constant $\tau_1$, which characterizes the rate of
polarization buildup, for transverse ($\bot$) and longitudinal ($||$)
orientation of beam and target polarization $Q$ is
\begin{eqnarray}
\tau_1^{\bot}=\frac{1}{\sigma_1 Q d_t f} \;\;\; \mathrm{and} \;\;\;
\tau_1^{||}=\frac{1}{(\sigma_1+\sigma_2) Q d_t f}\,
\label{eq:tau}
\end{eqnarray}
where $d_t$ is the target thickness in atoms/cm$^2$ and $f$ is the
revolution frequency of the particles in the ring. $\sigma_1$ and
$\sigma_2$ denote the spin--dependent total cross sections for
filtering with transverse and longitudinal target polarization. From
the measurement of the polarization buildup, the spin--dependent cross
sections can be determined. For small beam polarizations $P$, the
polarization buildup is linear in time. The spin--dependent cross
sections can be extracted from Eq.~(\ref{eq:tau}) using the known
target polarization, thickness, and the orbit frequency. In order to
extract both spin--dependent total cross sections, a measurement with
transverse and longitudinal beam polarization buildup is required. The
latter involves the operation of a Siberian snake in the AD.  It is
important to note that the buildup cross sections $\sigma_1$ and
$\sigma_2$, which we intend to measure as a function of the incident
beam energy and as a function of the ring acceptance angle, provide a
very convenient way to extract information about the spin--dependent
antiproton--proton interaction.

%%%%%%%%%%%%%%%%%%%%%%%%%%%%%%%%%%%%%%%%%%%%%%%%%%%%%%%%%%%%%%%%%%%%%%%%%%%%%
\section{Experimental Requirements for the AD--Ring \label{requirements}}
\pagestyle{myheadings} \markboth{Letter--of--Intent: Spin--Dependence of
$\bar{p}p$ Interaction at the AD}{Measurement Technique}
%%%%%%%%%%%%%%%%%%%%%%%%%%%%%%%%%%%%%%%%%%%%%%%%%%%%%%%%%%%%%%%%%%%%%%%%%%%%%
At present, the AD of CERN is actually the only place world wide,
where the proposed measurements can be performed. The effort involved
is substantial. Although we will perform most of the design work
outside of CERN, it is obvious, that many aspects in the design
require a close collaboration with the CERN machine group. The new
components that need to be installed in the AD are described in the
following sections. They shall all be tested and commissioned at the
Cooler Synchrotron COSY in J\"ulich. During these tests, we plan to
perform a few dedicated spin filtering experiments with protons.

\subsection{Low--$\beta$--Section}
The measurement requires implementing an internal polarized storage
cell target (PIT) in one of the straight sections of the AD. Targets
of this kind have been operated successfully at TSR in Heidelberg
\cite{filtex-target}, later on they were also used at HERA/DESY
\cite{hermes-target} at Indiana University Cyclotron Facility, and at
MIT--Bates. A new PIT is presently being commissioned at ANKE--COSY
\cite{SPIN}. A recent review can be found in
ref.~\cite{steffens-haeberli}. Typical target densities range from a
few 10$^{13}$ to $2\times10^{14}$~atoms/cm$^2$
\cite{hermes-target}. The target density depends strongly on the
transverse dimension of the storage cell. In order to provide a high
target density, the $\beta$--function at the storage cell should be
about $\beta_x=\beta_y=0.3$~m.  In order to minimize the
$\beta$--functions at the cell, a special insertion has to be
prepared, which includes additional quadrupoles around the storage
cell. The low--$\beta$ section should be designed in such a way that
the storage cell does not limit the machine acceptance. A careful
machine study has to be carried out in order to maintain the machine
performance at injection energy and at low energies for the other AD
experiments.
%A first machine study by Pavel
%Belochitsky from the AD machine group showed that  with the
%existing quadrupoles installed in the straight sections of the AD, at
%most values of $\beta_x=6$~m and $\beta_y=2$~m can be
%obtained. Although we will have to study this in more details, the
%experiment most likely requires to redesign the two straight sections
%and to install a new quadrupole magnet system. 
The section which houses the PIT has to be equipped with a powerful
differential pumping system, that is capable to maintain good vacuum
conditions in the other sections of the AD.

We will utilize the HERMES PIT (HERA/DESY), which will become
available at the beginning of 2006, to feed the storage cell. The
target will be operated in a weak magnetic guide field of a about
10~G. The orientation of the target polarization can be maintained by
a set of Helmholtz coils in transverse and longitudinal direction.

\subsection{Siberian Snake}
In order to determine $\sigma_2$, the stable beam spin direction has
to be longitudinal at the position of the PIT. Therefore, in the
straight section opposite the PIT, a solenoidal Siberian snake must be
implemented. A set of four skewed quadrupoles needs to be installed,
two before and two behind the snake to correct for the phase--space
rotation by the solenoid. We have begun to investigate whether
existing snakes can be utilized or modified to be used at the AD. In
any case, a careful machine study has to be carried out before final
conclusions can be reached.

\subsection{Electron Cooler}
The filtering experiments require to compensate multiple scattering in
the target by electron cooling. The present AD electron cooler is
capable to provide electron energies of up to 30~keV, corresponding to
antiproton beam energies of 50~MeV. In order to carry out the proposed
measurements in the energy range between 50 and 200 MeV, the electron
cooler at the AD should be upgraded to about 120~keV. Technically, this
solution seems feasible, whereas the installation of a new cooler,
such as the one previously installed at the TSL, involves major
modifications and re--commissioning of cooler and machine. At this
point, we believe that such an investment is not indicated, before we
have measured the spin--dependent cross sections at energies below 200
MeV.

\subsection{Intensity Increase in the AD through Stacking }
At present, the AD provides about $3\times 10^7$ stored
antiprotons. Through stacking, one may be able to increase the number
of stored antiprotons by about a factor of five, wherefrom the other
experiments at the AD would benefit as well. With a beam current
corresponding to about $10^8$ stored antiprotons, a luminosity of
$L=N_{\bar{p}} \cdot f \cdot d_t = 10^8 \cdot 10^6\,\mathrm{s^{-1}}
\cdot 10^{14} \, \mathrm{atoms/cm^2} = 10^{28}$~cm$^{-2}$s$^{-1}$ may
be achievable. For the purpose of polarimetry, this leads to elastic
antiproton--proton rates of several hundred events per second. To
achieve a larger number of stored antiprotons in the AD in the first
place is important, because after spin filtering for a few beam
lifetimes one wants to be left with a substantial beam intensity to
carry out beam polarization measurements. Once we have polarized the
beam, an unpolarized target can be used to determine the beam
polarization, thus the loss in beam intensity during filtering can be
compensated by an increase in target thickness.

%%%%%%%%%%%%%%%%%%%%%%%%%%%%%%%%%%%%%%%%%%%%%%%%%%%%%%%%%%%%%%%%%%%%%%%%%%%%%
\section{Polarimetry \label{polarimetry}}
\pagestyle{myheadings} \markboth{Letter--of--Intent: Spin--Dependence
of $\bar{p}p$ Interaction at the AD}{Polarimetry}
%%%%%%%%%%%%%%%%%%%%%%%%%%%%%%%%%%%%%%%%%%%%%%%%%%%%%%%%%%%%%%%%%%%%%%%%%%%%%
The experiments of the polarization buildup using stored antiprotons
should provide a measurement of the effective polarization buildup
cross section. The spin--dependent cross sections can be extracted
from the measured $dP/dt$, once the target polarization, the target
thickness, and the orbit frequency are known.  The target density can
be either obtained from the observed deceleration of the stored beam
when the electron cooling is switched off, as shown in
ref.~\cite{zapfe}, or it can be inferred from the measured rates in
the polarimeter using the quite well established elastic
antiproton--proton differential cross sections, measured at LEAR
\cite{klempt}. Thus an important subject is the development of a
polarimeter that allows one to efficiently determine the polarizations
of beam and target. Such a polarimeter based on silicon microstrip
detectors has recently been developed for the ANKE spectrometer
operated at the internal beam of COSY (\cite{schleichert}, more recent
information on the detection system can be found in
ref.~\cite{SPIN}). The use of this system as a polarimeter for our
experiments would neither require any additional R\&D, nor additional
costs.

There exist quite a number of analyzing power measurements for
antiproton--proton elastic scattering that can be employed
\cite{klempt}.  However, using the hydrogen PIT with an unpolarized
antiproton beam, it is possible to independently determine a suitable
polarization analyzer signal, which, when utilized in the analysis of
a polarized antiproton beam impinging on an unpolarized target,
provides through CPT invariance the polarization of the stored
antiproton beam.  The beam polarization achieved after spin filtering
in a longitudinally polarized target can be measured by switching off
adiabatically the Siberian snake, and subsequent left--right asymmetry
measurements. It is interesting to note that using the PIT, a direct
determination of the longitudinal spin correlation parameter $A_{zz}$
in elastic antiproton--proton scattering becomes possible.  Once this
parameter is established for the beam energies of interest, the
longitudinal beam polarization could be determined directly.

%%%%%%%%%%%%%%%%%%%%%%%%%%%%%%%%%%%%%%%%%%%%%%%%%%%%%%%%%%%%%%%%%%%%%%%%%%%%%
\section{Manpower and Cost Estimate, Timetable \label{manpowercost}}
\pagestyle{myheadings} \markboth{Letter--of--Intent: Spin--Dependence
of $\bar{p}p$ Interaction at the AD}{Manpower and Cost Estimate}
%%%%%%%%%%%%%%%%%%%%%%%%%%%%%%%%%%%%%%%%%%%%%%%%%%%%%%%%%%%%%%%%%%%%%%%%%%%%%
The present Letter--of--Intent is fully supported by the PAX
collaboration. It should be noted, that in all likelihood the amount
of work involved in setting up and running the proposed experiments at
the AD will not require all PAX collaborators. We are envisioning to
have available for the full proposal, which we plan to submit not
earlier than Summer 2006, a listing of the institutional
responsibilties for the AD experiment.

Below, we give an approximate timetable for the activities outlined
in this Letter--of--Intent. Prior to the installation, all components
will be tested off--site.
\begin{center}
\begin{tabular}{l|p{11cm}}
{\bf 2006--2007} &  Design and Construction Phase\\\hline
{\bf 2008}       &  Test of the low--$\beta$ target section, including the HERMES PIT and the Siberian Snake at COSY J\"ulich.\\ \hline
{\bf 2009}       & Installation of all components at the AD.\\\hline
{\bf 2009}       & 2 months of beam time at the AD, plus extra weeks of machine commissioning prior to the run.\\\hline
{\bf 2010}       & 2 months of beam time at the AD, plus extra weeks of machine commissioning prior to the run.
\end{tabular}
\end{center}
In Table~\ref{tab:work-estimate} the main components required for the
proposed studies are listed, as well as a distribution of work from
the side of the PAX collaboration and CERN. The estimated costs are
listed for those items that require constructive efforts.
%---------------------------------------------------------------------------------------------------
\begin{small}
\begin{table}[h!]
\begin{center}
\small\renewcommand{\arraystretch}{1}
\begin{tabular}{l|p{9cm}|r|r}
\hline 
\bf item    & \bf Component                             & {\bf Work}$^2$
                                                                                            & \bf Cost  \\
            &                                           & {\bf man--weeks}          & \bf kEuro        \\\hline\hline
1           & \bf Low--$\beta$ Section: 
               $\beta_x=\beta_y=0.3$~m                  &                                   &                    \\
1.1.1       & What is possible with small changes of 
              the AD-lattice                            &        1                          &                    \\
1.1.2       & Complete redesign of the Target Straight 
              Section                                   &        8                          &                    \\
1.2         & Lattice Study                             &        12                         &                    \\
1.3         & Tracking Study                            &        8                          &                    \\
1.4         & Documentation                             &        4                          &                    \\
1.5         & Inegration into the AD system             &        3                          &                    \\\hline\hline
2           & \bf Straight Sections                     &                                   &                    \\
2.1         & Documentation of the actual     
              System, Drawings etc.                     &        2                          &                    \\
2.2         & Detector Design                           &        10                         &                    \\
2.3         & Changes, space requirements               &        4                          &                    \\
2.4         & Slow control                              &        16                         & 120                \\
2.5         & Vacuum System Straight Sections           &        6                          & 500                \\
2.6         & Target Differential Pumping System        &        10                         & 500                \\
2.7         & Construction Off--Site                    &        12                         &                    \\
2.8         & 10 Quadrupoles low-$\beta$ Section        &                                   & 400                \\
2.9         & Electronics Experiment                    &                                   & 150                \\\hline\hline
3           & \bf Electron Cooler                       &                                   &                    \\        
3.1         & Upgrade existing Cooler to 120 keV        &        40                         & 200                \\
3.2         & Construction                              &        12                         &                    \\
3.3         & Commissioning                             &        12                         &                    \\
3.4         & Integration and Control                   &        16                         &                    \\ \hline\hline 
4           & \bf Siberian Snake                        &                                   &                    \\
4.1         & Snake Design                              &        4                          &                    \\
4.2         & Existing Snakes Decision Finding          &        6                          & 6                  \\
4.3         & Transport \& Hardware Tests               &        6                          & 7                  \\
4.4         & Lattice study for Implementation of Snake &        8                          &                    \\
4.5         & Integration into AD system                &       12                          &                    \\
4.6         & Power Supplies for Snake and 4 skewed 
              Quadrupoles                               &                                   & 100                \\\hline\hline
5           & \bf Beam Diagnostics                      &                                   &                    \\
5.1         & Design of near--target Beam Position 
              Monitors                                  &        6                          &                    \\
5.2         & Controls and low--level Electronics       &        4                          &                    \\
5.3         & Four Pickups with Electronics             &                                   & 120                \\\hline\hline
6           & \bf Construction On--Site: 12 weeks       &                                   &                    \\
6.1         & Target Section: 2 Engineers, 
                              2 Technicians, 
                              2 Workers                 &        72                         &                    \\
6.2         & Electron Cooler: 1 Engineer,
                               1 Technician,
                               2 Workers                &        48                         &                    \\
6.3         & Siberian Snake:  1 Engineer,
                               1 Technician,
                               2 Workers                &        48                         &                    \\\hline\hline
7           & \bf Commissioning: 12 weeks               &                                   &                    \\             
7.1         & 3 Engineers,
              3 Technicians,
              2 Workers                                 &        96                         &                    \\  
7.2         & Miscellaneous Electronic Material         &                                   & 60                 \\\hline\hline
            & Invest Total                              &                                   & 2163               \\
            & Travel Costs                              &                                   & 180                \\\hline\hline
\end{tabular}
\parbox{14cm}{\caption{\label{tab:work-estimate}\small List of
components, amount of work and cost estimates, required for the AD
Experiment.}}
\end{center}
\end{table} 
\end{small}
\footnote[2]{CERN help will be required for the preparation of the
experiment, the amount will be evaluated after a more detailed study
is available.}
%---------------------------------------------------------------------------------------------------
%%%%%%%%%%%%%%%%%%%%%%%%%%%%%%%%%%%%%%%%%%%%%%%%%%%%%%%%%%%%%%%%%%%%%%%%%%%%%
\pagestyle{myheadings} \markboth{Letter--of--Intent: Spin--Dependence
of $\bar{p}p$ Interaction at the AD}{Acknowledgement}
%%%%%%%%%%%%%%%%%%%%%%%%%%%%%%%%%%%%%%%%%%%%%%%%%%%%%%%%%%%%%%%%%%%%%%%%%%%%%
\section{Acknowledgement}
We would like to thank Pavel Belochitskii, Stephan Maury, Dieter
M\"ohl, Gerard Tranquille, and Tommy Eriksson of the CERN machine
staff for their support and the many helpful suggestions brought up
during our site visits.
%%%%%%%%%%%%%%%%%%%%%%%%%%%%%%%%%%%%%%%%%%%%%%%%%%%%%%%%%%%%%%%%%%%%%%%%%%%%%%%%%%%%%
\pagestyle{myheadings} \markboth{Letter--of--Intent: Spin--Dependence of
$\bar{p}p$ Interaction at the AD}{Bibliography}
%%%%%%%%%%%%%%%%%%%%%%%%%%%%%%%%%%%%%%%%%%%%%%%%%%%%%%%%%%%%%%%%%%%%%%%%%%%%%%%%%%%%%

%%%%%%%%%%%%%%%%%%%%%%%%%%%%%%%%%%%%%%%%%%%%%%%%%%%%%%%%%%%%%%%%%%%%%%%%%%%%%%%%%%%%%
\cleardoublepage
\section{Appendix A} 
\pagestyle{myheadings} \markboth{Letter--of--Intent: Spin--Dependence of
$\bar{p}p$ Interaction at the AD}{Appendix A: PAX Collaboration List}
\begin{appendix}
%%%%%%%%%%%%%%%%%%%%%%%%%%%%%%%%%%%%%%%%%%%%%%%%%%%%%%%%%%%%%%%%%%%%%%%%%%%%%
\section*{Members of the Collaboration}
%%%%%%%%%%%%%%%%%%%%%%%%%%%%%%%%%%%%%%%%%%%%%%%%%%%%%%%%%%%%%%%%%%%%%%%%%%%%%
\subsection*{Alessandria, Italy, Universita$'$ del Piemonte 
Orientale $''$A. Avogadro$''$ and INFN} 
Vincenzo Barone

\subsection*{Beijing, China, School of Physics, Peking University}
Bo--Qiang Ma    

\subsection*{Bochum, Germany, Institut f\"ur 
Theoretische Physik II, Ruhr Universit\"at Bochum}
Klaus Goeke, 
Andreas Metz, and
Peter Schweitzer

\subsection*{Bonn, Germany, Helmholtz--Institut f\"ur Strahlen-- 
und Kernphysik, Universit\"at Bonn}
Paul--Dieter Eversheim, 
Frank Hinterberger, 
Ulf--G. Mei{\ss}ner,
Heiko Rohd\-je{\ss}, and
Alexander Sibirtsev

\subsection*{Brookhaven, USA, Collider--Accelerator Department, Brookhaven National Laboratory}
Christoph Montag

\subsection*{Brookhaven, USA, RIKEN BNL Research Center, Brookhaven National Laboratory}
Werner Vogelsang

\subsection*{Cagliari, Italy, Dipartimento di Fisica, Universita$'$ di Cagliari and INFN}
Umberto D$'$Alesio, and 
Francesco Murgia

\subsection*{Dublin, Ireland, School  of Mathematics, Trinity College, University of Dublin}
Nigel Buttimore

\subsection*{Dubna, Russia, Joint Institute for Nuclear Research} 
Sergey Dymov,
Anatoly Efremov, 
Oleg Ivanov,              
Natela Kadagidze, 
Vladimir Komarov, 
Victor Krivokhizhin,
Anatoly Kulikov,
Vladimir Kurbatov,
Vladimir Leontiev,
Gogi Macharashvili,
Sergey Merzliakov,
Gleb Meshcheryakov,
Igor Meshkov,
Alexander Nagaytsev,
Vladimir Peshekhonov, 
Igor Savin, 
Valeri Serdjuk,
Binur Shaikhatdenov,
Oleg Shevchenko, 
Anatoly Sidorin,
Alexander Smirnow,
Evgeny Syresin,
Oleg Teryaev,
Sergey Trusov, 
Yuri Uzikov,
Gennady Yarygin,
Alexander Volkov, and
Nikolai Zhuravlev

\subsection*{Erlangen, Germany, Physikalisches Institut, Universit\"at 
Er\-langen--N\"urn\-berg} 
Wolfgang Eyrich,
Andro Kacharava,
Bernhard Krauss,
Albert Lehmann,
Alexander Nass,
Davide Reggiani,
Klaus Rith,
Ralf Seidel,
Erhard Steffens,
Friedrich Stinzing,
Phil Tait, and
Sergey Yaschenko

\subsection*{Ferrara, Italy, Istituto Nazionale di Fisica Nucleare}
Marco Capiluppi, 
Guiseppe Ciullo, 
Marco Contalbrigo, 
Alessandro Drago, 
Paola Ferretti--Dalpiaz, 
Francesca Giordano,
Paolo Lenisa, 
Luciano Pappalardo,
Giulio Stancari,
Michelle Stancari, and
Marco Statera

\subsection*{Frascati, Italy, Istituto Nazionale di Fisica Nucleare}
Eduard Avetisyan,
Nicola Bianchi,
Enzo De Sanctis,
Pasquale Di Nezza,
Alessandra Fantoni,
Cynthia Hadjidakis,
Delia Hasch,
Marco Mirazita,
Valeria Muccifora,
Federico Ronchetti, and
Patrizia Rossi

\subsection*{Gatchina, Russia, Petersburg Nuclear Physics Institute}
Sergey Barsov,
Stanislav Belostotski, 
Oleg Grebenyuk, 
Kirill Grigoriev,
Anton Izotov, 
Anton Jgoun,
Peter Kravtsov, 
Sergey Manaenkov, 
Maxim Mikirtytchiants,
Sergey Mikirtytchiants,
Oleg Miklukho,  
Yuri Naryshkin, 
Alexander Vassiliev, and 
Andrey Zhdanov 

\subsection*{Gent, Belgium,  Department of Subatomic and 
Radiation Physics, University of Gent} 
Dirk Ryckbosch

\subsection*{Hefei, China, Department of Modern Physics, University of Science and Technology
of China}
Yi Jiang,
Hai--jiang Lu,
Wen--gan Ma,
Ji Shen,
Yun--xiu Ye,
Ze--Jie Yin, and
Yong--min Zhang

\subsection*{J\"ulich, Germany, Forschungszentrum J\"ulich, Institut 
f\"ur Kernphysik} 
David  Chiladze, 
Ralf Gebel,
Ralf Engels, 
Olaf Felden,
Johann Haidenbauer,
Christoph Hanhart,  
Michael Hartmann,
Irakli Keshelashvili,
Siegfried Krewald,
Andreas Lehrach,   
Bernd Lorentz,  
Sigfried Martin, 
Ulf--G. Mei{\ss}ner,
Nikolai Nikolaev,  
Dieter Prasuhn, 
Frank Rathmann, 
Ralf Schleichert,
Hellmut Seyfarth,  and
Hans Str\"oher

\subsection*{Kosice, Slovakia, Institute of Experimental Physics, 
Slovak Academy of Sciences and P.J. Safarik University, Faculty of Science} 
Dusan Bruncko, 
Jozef Ferencei, 
J\'an Mu\v sinsk\'y, and  
Jozef Urb\'an

\subsection*{Langenbernsdorf, Germany, Unternehmensberatung und Service--B\"uro (USB), 
Gerlinde Schulteis \& Partner GbR} 
Christian Wiedner (formerly at MPI-K Heidelberg)

\subsection*{Lecce, Italy, Dipartimento di Fisica, Universita$'$ di Lecce and INFN}
Claudio Corian\'o, and 
Marco Guzzi 

\subsection*{Madison, USA, University of Wisconsin} 
Tom Wise

\subsection*{Milano, Italy, Universita' dell'Insubria, Como and INFN sez.} 
Philip Ratcliffe

\subsection*{Moscow, Russia, Institute for Theoretical and 
Experimental Physics}
Vadim Baru,
Ashot Gasparyan,  
Vera Grishina, 
Leonid Kondratyuk, and
Alexander Kudriavtsev

\subsection*{Moscow, Russia, Lebedev Physical Institute}
Alexander Bagulya,
Evgeni Devitsin,
Valentin Kozlov,
Adel Terkulov, and
Mikhail Zavertiaev

\subsection*{Moscow, Russia, Physics Department, Moscow Engineering Physics Institute}
Aleksei Bogdanov,
Sandibek Nurushev,
Vitalii Okorokov,
Mikhail Runtzo, and
Mikhail Strikhanov

\subsection*{Novosibirsk, Russia, Budker Institute for Nuclear Physics}
Yuri Shatunov

\subsection*{Palaiseau, France, Centre de Physique Theorique, Ecole Polytechnique} 
Bernard Pire

\subsection*{Protvino, Russia, Institute of High Energy Physics} 
Nikolai Belikov,     
Boris Chujko,     
Yuri Kharlov,    
Vladislav Korotkov,   
Viktor Medvedev,    
Anatoli Mysnik,     
Aleksey Prudkoglyad,
Pavel Semenov,   
Sergey Troshin, and
Mikhail Ukhanov 

\subsection*{Tbilisi, Georgia, Institute of High Energy Physics and Informatization, 
Tbilisi State University} 
Badri Chiladze,
Archil Garishvili,
Nodar Lomidze,
Alexander Machavariani,
Mikheil Nioradze,
Tariel Sakhelashvili, 
Mirian Tabidze, and
Igor Trekov

\subsection*{Tbilisi, Georgia, Nuclear Physics Department, Tbilisi State University} 
Leri Kurdadze, and 
George Tsirekidze

\subsection*{Torino, Italy, Dipartimento di Fisica Teorica, 
Universita di Torino and INFN} 
Mauro Anselmino, 
Mariaelena Boglione, and
Alexei Prokudin

\subsection*{Uppsala, Sweden, Department of Radiation Sciences, Nuclear Physics Division} 
Pia Thorngren--Engblom

\subsection*{Virginia, USA, Department of Physics, University of Virginia} 
Simonetta Liuti

\subsection*{Warsaw, Poland, Soltan Institute for Nuclear Studies} 
Witold Augustyniak, 
Bohdan Marianski, 
Lech Szymanowski, 
Andrzej Trzcinski, and
Pawel Zupranski

\subsection*{Yerevan, Armenia, Yerevan Physics Institute}
Norayr Akopov, 
Robert Avagyan,
Albert Avetisyan,
Garry Elbakyan,
Zaven Hakopov,
Hrachya Marukyan, and
Sargis Taroian

\end{appendix}
\end{document}